\documentclass[referee]{raa}

\usepackage{colortbl}
\usepackage{graphicx,times}
\usepackage{natbib}
\usepackage{makecell}
\usepackage{mathtools}
\usepackage{multirow}
\usepackage{amssymb,amsmath}
\usepackage{booktabs}
\usepackage{physics}
\usepackage{url}

\bibpunct{(}{)}{;}{a}{}{,}

\usepackage[colorlinks,linkcolor=blue,anchorcolor=blue,citecolor=blue,
urlcolor=blue]{hyperref}

\usepackage{tikz}
\usetikzlibrary{shapes, arrows.meta, positioning, fit, backgrounds,
shadows, calc}
\usepackage[most]{tcolorbox}

\newtcolorbox{agentlog}{
    breakable,
    enhanced,
    colback=gray!5,
    colframe=gray!40,
    boxrule=0.5pt,
    arc=2pt,
    left=4pt, right=4pt, top=4pt, bottom=4pt,
    fontupper=\small\sffamily
}

\newcommand{\internal}[1]{%
    \par\noindent
    \parbox{\dimexpr\linewidth-1em}{
        \footnotesize\sffamily\color{gray!80} #1
    }\par
}

\newcommand{\role}[1]{\textbf{#1}}
\newcommand{\action}[2]{\par\noindent\texttt{\textcolor{blue!40!gray}{[#1]}} #2}
\newcommand{\annotation}[1]{\par\noindent\textit{\textcolor{gray!60}{(#1)}}}

\begin{document}

\title{DeepInflation: an AI agent for research and model discovery of inflation}

\volnopage{Vol.0 (202x) No.0, 000--000}
\setcounter{page}{1}

\author{Ze-Yu Peng
    \inst{1,2,3}
    \and Hao-Shi Yuan
    \inst{1,4,5}
    \and Qi Lai
    \inst{1,4,5}
    \and Qing-Yu Lan
    \inst{1}
    \and Zhan-He Wang
    \inst{1}
    \and Jun-Qian Jiang
    \inst{1}
    \and Gen Ye
    \inst{6,1}
    \and Jun Zhang
    \inst{2,3}
    \and Yun-Song Piao
    \inst{1,2,4,5}
}

\institute{School of Physical Sciences, University of Chinese Academy of Sciences, Beijing 100049, China; \email{pengzeyu23@mails.ucas.ac.cn}\\
    \and
    International Centre for Theoretical Physics Asia-Pacific, University of Chinese Academy of Sciences, 100190 Beijing, China\\
    \and
    Taiji Laboratory for Gravitational Wave Universe, University of Chinese Academy of Sciences, 100049 Beijing, China\\
    \and
    School of Fundamental Physics and Mathematical Sciences, Hangzhou Institute for Advanced Study, UCAS, Hangzhou 310024, China\\
    \and
    Institute of Theoretical Physics, Chinese Academy of Sciences, P.O. Box 2735, Beijing 100190, China\\
    \and
    D\'{e}partement de Physique Th\'{e}orique, Universit\'{e} de Gen\`{e}ve, 24 quai Ernest-Ansermet, CH-1211 Gen\`{e}ve 4, Switzerland\\
    \vs\no
}

\abstract{We present \textbf{DeepInflation}, an AI agent designed for
    research and model discovery in inflationary cosmology. Built upon
    a multi-agent architecture, \textbf{DeepInflation} integrates
    Large Language Models (LLMs) with a symbolic regression (SR)
    engine and a retrieval-augmented generation (RAG) knowledge base.
    This framework enables the agent to automatically explore and
    verify the vast landscape of inflationary potentials while
    grounding its outputs in established theoretical literature. We
    demonstrate that \textbf{DeepInflation} can successfully discover
    simple and viable single-field slow-roll inflationary potentials
    consistent with the latest observations (with the ACT DR6 results taken as an example)
    or any given $n_s$ and $r$,
    and provide accurate theoretical context for obscure inflationary
    scenarios. \textbf{DeepInflation} serves as a prototype for a new
    generation of autonomous scientific discovery engines in
    cosmology, which enables researchers and non-experts alike to
    explore the inflationary landscape using natural language. This
    agent is available at
    \url{https://github.com/pengzy-cosmo/DeepInflation}.
}

\maketitle

\section{Introduction}\label{sec:intro}

Cosmic inflation
(\citealt{Guth:1980zm,Linde:1981mu,Albrecht:1982wi,Starobinsky:1980te,Mukhanov:1981xt}),
the standard paradigm of the very early universe, has become a
cornerstone of modern cosmology. While resolving several
fundamental problems in standard Big Bang cosmology, it provides a
natural mechanism for generating the primordial perturbations that
are in excellent agreement with observations of the cosmic
microwave background (CMB) (\citealt{Planck:2018vyg}).

Despite the success of the inflationary paradigm, the underlying
physics driving inflation remains unknown. This theoretical
uncertainty has led to a remarkably diverse landscape of
inflationary models, even within the simplest single-field
slow-roll scenario (\citealt{Martin:2013tda,Kallosh:2025ijd}). As
cosmological observations become increasingly precise
(\citealt{WMAP:2012nax,Planck:2018vyg,AtacamaCosmologyTelescope:2025blo,
AtacamaCosmologyTelescope:2025nti,SPT-3G:2025bzu}), the constraints
on inflationary parameters are continuously tightening. Notably,
the latest ACT DR6 (\citealt{AtacamaCosmologyTelescope:2025blo,
AtacamaCosmologyTelescope:2025nti}) prefers a higher scalar
spectral index $n_s$ compared to Planck, which rules out the
previously favored Starobinsky model
(\citealt{Starobinsky:1980te,Starobinsky:1992ts}) at more than
$2\sigma$, see also recent (\citealt{Linde:2025pvj}). This result has
prompted extensive interest in re-evaluating existing models and
proposing new theories to reconcile with the latest data
(\citealt{Kallosh:2025rni,
Berera:2025vsu,Dioguardi:2025vci,Brahma:2025dio,Gialamas:2025kef,Liu:2025qca,Gao:2025onc,Gialamas:2025ofz,Heidarian:2025drk,Odintsov:2025eiv,Bezerra-Sobrinho:2025gfg,Fu:2025ciy,Qiu:2025uot,Aoki:2025wld,Aoki:2025ywt}).
In addition, constraints on primordial observables are also
sensitive to the cosmological tensions that $\Lambda$CDM is
suffering from
(\citealt{Ye:2021nej,Jiang:2023bsz,Ferreira:2025lrd,McDonough:2025lzo}).
It is found that the proposed resolutions of the Hubble tension
often lead to a scale invariant spectrum with $n_s \approx 1$
(\citealt{Ye:2020btb,Jiang:2022uyg,Jiang:2022qlj,Smith:2022hwi,Peng:2023bik,Wang:2024tjd,Peng:2025tqt}).
Therefore, efficient tools for exploring the vast model space and
confronting theoretical predictions with observational data are
crucial.

The rapid advancement of large language models (LLMs) in recent
years opens new avenues for scientific research
(\citealt{zhao2025surveylargelanguagemodels,
        luo2025llm4srsurveylargelanguage, tie2025surveyaiscientists,
Barman_2025}). LLMs can effectively process and generate human-like
text, making them valuable tools for literature review, hypothesis
generation, and even manuscript drafting. However, standalone LLMs
face inherent limitations when applied to rigorous scientific
tasks. They are prone to hallucinations, generating plausible but
incorrect statements, especially when lacking domain-specific
training~(\citealt{Huang_2025}). Furthermore, LLMs lack the capability
to perform precise numerical calculations, such as solving
differential equations, which are essential for accurate model
analysis~(\citealt{meadows2024exploringlimitsfinegrainedllmbased}).

The emergence of LLM-based AI agents offers a promising solution
to these challenges. Unlike passive chatbots, an agent acts as a
reasoning engine that can autonomously plan tasks and utilize
external tools (\citealt{yao2023reactsynergizingreasoningacting}). By
integrating LLMs with domain-specific knowledge bases and rigorous
numerical solvers, agents can ground their natural language
reasoning in verified physics and precise calculations. Recent
works have demonstrated the potential of such systems in
cosmological research, including CMBAgent
(\citealt{Laverick:2024fyh,xu2025opensourceplanning}) for automated
cosmological parameter analysis, CLAPP (\citealt{Casas:2025hcz}) for
code assistance, and SimAgents (\citealt{Zhang:2025uoo}) for extracting
simulation parameters from literature.

Symbolic regression (SR) (\citealt{schmidt2009distilling,cranmer2023})
has emerged as a powerful technique for automated scientific
discovery. By searching for explicit analytical expressions that
optimize a given objective, SR enables the discovery of
interpretable models directly from data without relying on
predefined functional forms or black-box approximators. It has
been recently considered that the combination of symbolic
regression with LLMs represents a nascent but highly promising
avenue for future development
(\citealt{bartlett2025introductionsymbolicregressionphysical}).

In this work, we introduce \textbf{DeepInflation}, an AI agent
specifically designed for inflationary cosmology research and
model discovery. Based on a multi-agent architecture,
\textbf{DeepInflation} integrates a SR engine for automated model
discovery and a domain-specific Retrieval-Augmented Generation
(RAG)
system~(\citealt{lewis2021retrievalaugmentedgenerationknowledgeintensivenlp})
constructed from the \textit{Encyclop\ae dia Inflationaris}
(\citealt{Martin:2013tda}). It enables researchers and non-experts
alike to explore the inflationary landscape using natural
language, thereby facilitating the efficient discovery and
verification of inflationary models that align with the latest
observational constraints.

The rest of this paper is organized as follows. In
Sect.~\ref{sec:architecture}, we describe the system architecture
and the design of each component. Sect.~\ref{sec:examples}
demonstrates the agent's capabilities through representative use
cases. We conclude with a discussion of limitations and an outlook
in Sect.~\ref{sec:discussion}.

\section{Architecture}\label{sec:architecture}

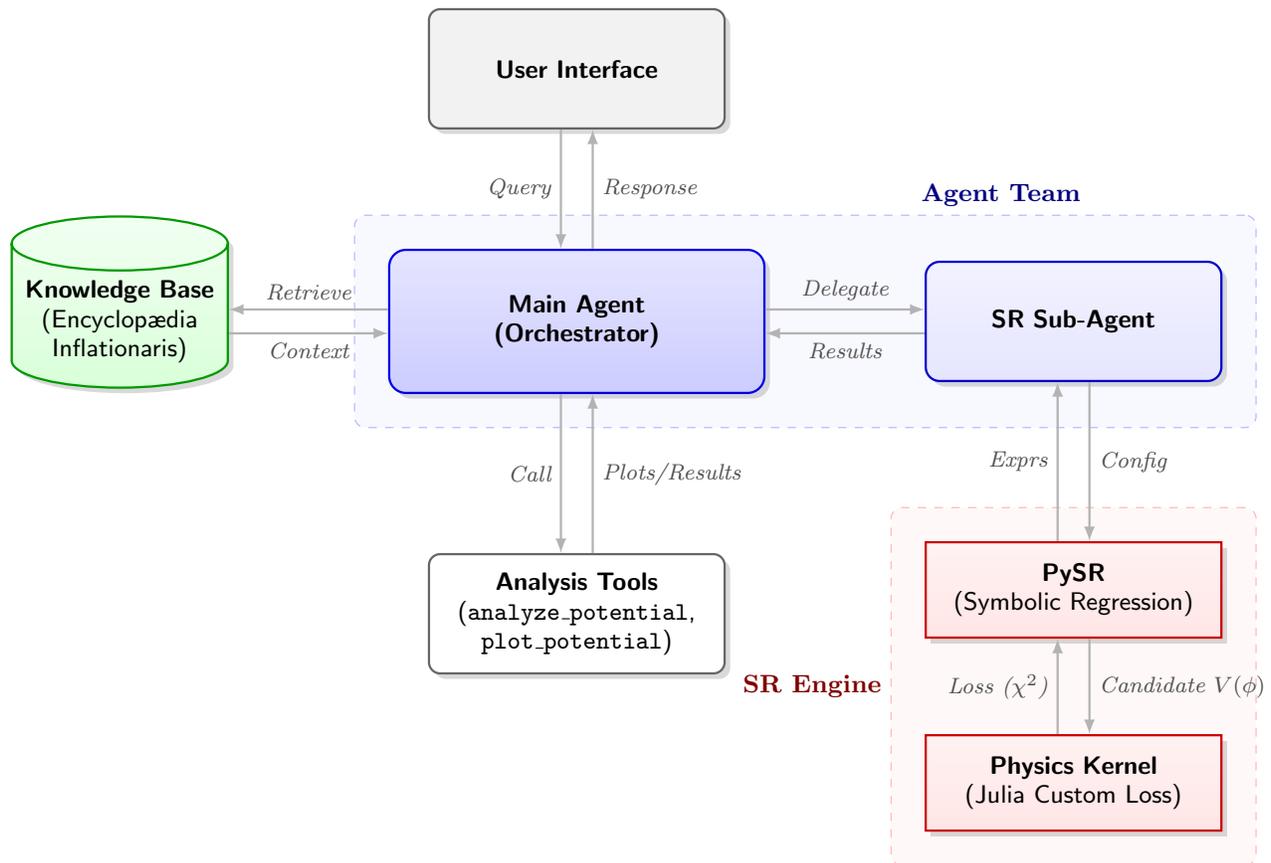
\begin{figure*}[htbp]
    \centering
    \resizebox{0.95\textwidth}{!}{
        \begin{tikzpicture}[
                node distance=1.5cm and 2cm,
                >=latex,
                font=\sffamily\small,
                base_node/.style={
                    align=center,
                    drop shadow={opacity=0.3, shadow xshift=2pt, shadow yshift=-2pt},
                    thick
                },
                block/.style={
                    base_node,
                    rectangle,
                    rounded corners,
                    draw=gray!80!black,
                    fill=white,
                    text width=3.5cm,
                    minimum height=1.5cm,
                },
                agent/.style={
                    base_node,
                    rectangle,
                    rounded corners=5pt,
                    draw=blue!80!black,
                    top color=blue!5,
                    bottom color=blue!10,
                    text width=3.5cm,
                    minimum height=1.5cm
                },
                main_agent_style/.style={
                    base_node,
                    rectangle,
                    rounded corners=6pt,
                    draw=blue!90!black,
                    top color=blue!10,
                    bottom color=blue!20,
                    text width=4.5cm,
                    minimum height=1.8cm,
                    font=\sffamily\bfseries
                },
                engine/.style={
                    base_node,
                    rectangle,
                    draw=red!80!black,
                    top color=red!5,
                    bottom color=red!10,
                    text width=3.5cm,
                    minimum height=1.2cm
                },
                database/.style={
                    base_node,
                    cylinder,
                    shape border rotate=90,
                    draw=green!60!black,
                    top color=green!5,
                    bottom color=green!15,
                    aspect=0.25,
                    text width=2.5cm,
                    minimum height=1.5cm
                },
                label_text/.style={font=\footnotesize\itshape, text=black!70},
                group_box/.style={
                    draw=gray!40,
                    dashed,
                    inner sep=12pt,
                    rounded corners,
                    fill=gray!5,
                    fill opacity=0.5
                }
            ]

            \node[block, fill=gray!10] (user) {\textbf{User Interface}};

            \node[main_agent_style, below=1.5cm of user] (main_agent)
            {\textbf{Main Agent}\\ (Orchestrator)};

            \node[database, left=2cm of main_agent] (rag)
            {\textbf{Knowledge Base}\\ (Encyclop\ae dia Inflationaris)};

            \node[agent, right=2cm of main_agent] (sr_agent) {\textbf{SR Sub-Agent}};

            \node[block, below=2cm of main_agent] (tools) {\textbf{Analysis
            Tools}\\ (\texttt{analyze\_potential}, \texttt{plot\_potential})};

            \node[engine, below=2cm of sr_agent] (pysr)
            {\textbf{PySR}\\ (Symbolic Regression)};

            \node[engine, below=1.2cm of pysr] (julia) {\textbf{Physics
            Kernel}\\ (Julia Custom Loss)};


            \draw[->, thick, gray!60] ([xshift=-2mm]user.south) --
            node[left, label_text] {Query} ([xshift=-2mm]main_agent.north);
            \draw[->, thick, gray!60] ([xshift=2mm]main_agent.north) --
            node[right, label_text] {Response} ([xshift=2mm]user.south);

            \draw[->, thick, gray!60] ([yshift=1.5mm]main_agent.west) --
            node[above, label_text] {Retrieve} ([yshift=1.5mm]rag.east);
            \draw[->, thick, gray!60] ([yshift=-1.5mm]rag.east) --
            node[below, label_text] {Context} ([yshift=-1.5mm]main_agent.west);

            \draw[->, thick, gray!60] ([yshift=1.5mm]main_agent.east) --
            node[above, label_text] {Delegate} ([yshift=1.5mm]sr_agent.west);
            \draw[->, thick, gray!60] ([yshift=-1.5mm]sr_agent.west) --
            node[below, label_text] {Results} ([yshift=-1.5mm]main_agent.east);

            \draw[->, thick, gray!60] ([xshift=-2mm]main_agent.south) --
            node[left, label_text] {Call} ([xshift=-2mm]tools.north);
            \draw[->, thick, gray!60] ([xshift=2mm]tools.north) --
            node[right, label_text] {Plots/Results} ([xshift=2mm]main_agent.south);

            \draw[->, thick, gray!60] ([xshift=2mm]sr_agent.south) --
            node[right, label_text] {Config} ([xshift=2mm]pysr.north);
            \draw[->, thick, gray!60] ([xshift=-2mm]pysr.north) --
            node[left, label_text] {Exprs} ([xshift=-2mm]sr_agent.south);

            \draw[->, thick, gray!60] ([xshift=2mm]pysr.south) --
            node[right, label_text] {Candidate $V(\phi)$} ([xshift=2mm]julia.north);
            \draw[->, thick, gray!60] ([xshift=-2mm]julia.north) --
            node[left, label_text] {Loss ($\chi^2$)} ([xshift=-2mm]pysr.south);

            \begin{pgfonlayer}{background}

                \node[group_box, draw=blue!30, fill=blue!5, fit=(main_agent)
                    (sr_agent), label={[blue!50!black, font=\bfseries]above
                right:Agent Team}] (agent_group) {};

                \node[group_box, draw=red!30, fill=red!5, fit=(pysr) (julia),
                label={[red!50!black, font=\bfseries]west:SR Engine}]
                (sr_engine_group) {};
            \end{pgfonlayer}

        \end{tikzpicture}
    }
    \caption{System architecture of \textbf{DeepInflation}. The system
        is composed of an \textbf{Agent Team}  and a \textbf{SR
        engine}. The Main Agent orchestrates the workflow, retrieving
        theoretical context from the Knowledge Base and delegating model
        discovery tasks to the SR Sub-Agent. The SR engine employs PySR
        for genetic search, coupled with a Julia-based Physics Kernel that
        evaluates the $\chi^2$ loss by solving the inflationary equations
    of motion.}
    \label{fig:architecture}
\end{figure*}

The architecture of \textbf{DeepInflation} is illustrated in
Fig.~\ref{fig:architecture}. We construct a
multi-agent system, which comprises a main orchestrating agent and
a specialized sub-agent for SR, coordinated through a delegation
mechanism. The main agent has access to three core capabilities:
(i) a SR engine for model discovery, (ii) a domain-specific RAG
knowledge base constructed from the \textit{Encyclop\ae dia
Inflationaris}, and (iii) analysis and visualization tools for
computing and presenting inflationary observables. Powered by
advanced LLMs, the main agent interprets user queries, plans
workflows, and automatically invokes the appropriate tools and
sub-agents to implement the tasks.

\subsection{Symbolic Regression}\label{subsec:sr}

The primary capability of \textbf{DeepInflation} is the automated
discovery of inflationary potentials that satisfy specific
observational constraints. This is achieved through symbolic
regression (SR), a machine learning technique that searches for
explicit analytical expressions optimizing a given objective.
Previous studies have demonstrated the potential of SR in exploring inflationary models (\citealt{Kamerkar:2022dfu,Sousa:2023unz}).
We employ PySR~(\citealt{cranmer2023}), a high-performance SR library,
as the search backend,
and introduce a SR sub-agent that interprets the user requirements
into appropriate constraints and settings for SR tasks, and
autonomously executes the search process.

The discovery of inflationary models requires finding a potential
$V(\phi)$, where $\phi$ is the inflaton, that yields observables
consistent with target constraints. We achieve this by
implementing a custom loss function that integrates the equations
of motion for single-field slow-roll inflation, see details in
e.g.~(\citealt{Baumann:2022mni}). The loss function for each
candidate potential $V(\phi)$ is computed through the following
steps:

\begin{enumerate}
    \item \textbf{Inflation}: The end of inflation $\phi_{\text{end}}$
        is determined by solving $\epsilon(\phi_{\text{end}}) = 1$, where
        $\epsilon = \frac{M_{\rm Pl}^2}{2}(V'/V)^2$ is the slow-roll
        parameter of inflation. The trajectory is then integrated
        backwards by solving the equation:
        \begin{equation}
            \int_{\phi_{\text{end}}}^{\phi_N}{d\phi} = \int \left(M_{\rm
            Pl}^2\frac{ V'}{V}\right)dN,
        \end{equation}
        to find the horizon-crossing field value $\phi_N$ corresponding to
        $N$ e-folds before the end of inflation.
    \item \textbf{Observables}: The scalar spectral index $n_s$ and
        tensor-to-scalar ratio $r$ are computed using the slow-roll
        approximations ($\epsilon\ll1$, $\eta = M_{\rm Pl}^2 V''/V\ll 1$)
        at $\phi_N$:
        \begin{equation}
            n_s = 1 - 6\epsilon_N + 2\eta_N, \quad r = 16\epsilon_N.
        \end{equation}
    \item \textbf{Custom Loss Function}: The fitness of the candidate
        potential is then evaluated using the $\chi^2$ distance in the
        $(n_s, r)$ plane based on the target observational constraints:
        \begin{equation}\label{eq:loss}
            2\mathcal{L} = \left( \frac{n_s - n_s^{\rm obs}}{\sigma_{n_s}}
            \right)^2 + \left( \frac{r - r^{\rm obs}}{\sigma_r} \right)^2,
        \end{equation}
        where $n_s^{\rm obs}$ and $r^{\rm obs}$ are the target values
        instructed by the user, and $\sigma_{n_s}$, $\sigma_r$ are the
        corresponding uncertainties.
\end{enumerate}
Note that $A_s$ is not included in the loss function, as it
constrains only the overall energy scale of the potential and does
not affect its shape; the discovered expressions are therefore
determined up to a normalization factor.
To integrate this loss function with PySR, we implement the
inflationary calculations in Julia, which also provides high
efficiency for numerical calculations.

Once the SR sub-agent receives the results from the SR engine, it
summarizes the best potentials based on their loss values and
simplicity, and then returns them to the main agent for final
response output.

\subsection{RAG Knowledge Base}\label{subsec:rag}

To ground the agent's responses in established domain knowledge,
we construct a RAG
system~(\citealt{lewis2021retrievalaugmentedgenerationknowledgeintensivenlp})
from the \textit{Encyclop\ae dia Inflationaris} (ASPIC)
(\citealt{Martin:2013tda}). This comprehensive library contains over 70
distinct inflationary scenarios and their corresponding physical
properties.

We employ a hybrid search strategy combining dense vector retrieval
with sparse keyword search (BM25), which enhances the accuracy of
physical noun retrieval compared to the standard semantic search alone.
The source text is chunked by model sections, ensuring that the
retrieved context preserves the complete theoretical description of a
model. This allows the agent to accurately answer queries about
specific models, retrieve their potential forms, and understand their
theoretical motivations in detail.

\subsection{Analysis and Visualization Tools}\label{subsec:tools}

Beyond model discovery, \textbf{DeepInflation} provides tools for
analysis and visualization of models.
\begin{itemize}
    \item \texttt{analyze\_potential}: Computes observables for a given
        potential expression. This tool follows the same procedure as the
        loss function in the SR engine, allowing the agent to verify the
        predictions of discovered models or user-specified potentials.
    \item \texttt{plot\_potential}: Generates diagnostic figures for a
        given potential, including the shape of the potential $V(\phi)$,
        the evolution of slow-roll parameters $\epsilon(\phi)$ and
        $\eta(\phi)$, and the model's prediction in the $n_s - r$ plane
        against the observational contours (Planck+BICEP/Keck
        2018)~(\citealt{BICEP:2021xfz}).
\end{itemize}
These tools are exposed to the main agent as function calls, enabling
a workflow where the agent can ``discover'' a candidate model via SR,
``verify'' it using the analyzer, and ``visualize'' the result for the user.

\section{Examples}\label{sec:examples}

To demonstrate the capabilities of \textbf{DeepInflation}, we
present two representative use cases: (i) discovering inflationary
models consistent with the latest CMB observations, and (ii)
inquiring about specific inflationary models. In our examples, we
use the latest LLM model from OpenAI, GPT-5.2, with temperature
set to $1.0$. The embedding model for RAG is
\texttt{text-embedding-3-small} from OpenAI.

\subsection{Example: Model Discovery (according to ACT DR6 results)}

The ACT DR6 data favors a scalar spectral index $n_s$ higher than
the Planck 2018 result, which disfavors the Starobinsky model
(\citealt{AtacamaCosmologyTelescope:2025blo,
AtacamaCosmologyTelescope:2025nti}). We task the agent to find a
simple inflationary model consistent with these new constraints.

\begin{figure*}[htbp]
    \centering
    \includegraphics[width=0.95\linewidth]{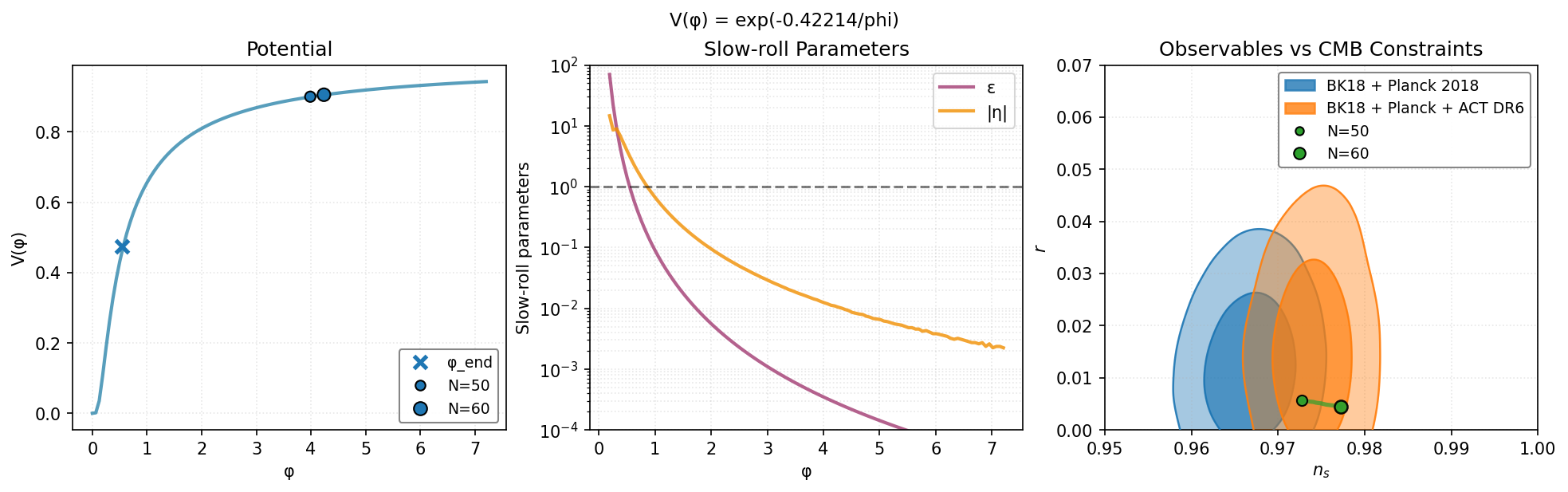}
    \caption{Diagnostic plot generated by the agent for the discovered
        best model $V(\phi)=\exp\!\left(-\frac{0.42214}{\phi}\right)$.
        The right panel shows the predictions in the $n_s$--$r$ plane
        overlaid on the BK18+Planck 2018 and BK18+Planck+ACT DR6
    observational contours~(\citealt{BICEP:2021xfz,AtacamaCosmologyTelescope:2025blo}).}
    \label{fig:case_study_plot}
\end{figure*}

The main agent delegates the task to the SR sub-agent, which then
configures and runs the SR engine with appropriate physics
targets. The main agent receives the search results and outputs
the final response to the user. We then ask the agent to generate
a diagnostic plot for this model, which is shown in
Fig.~\ref{fig:case_study_plot}.

\begin{agentlog}
    \role{User}: I want to find a simple inflation model that matches
    the latest CMB observations, with $n_s = 0.974 \pm 0.003$.

    \vspace{0.5em}
    \internal{
        \role{Main Agent}:
        \action{Delegate to SR Sub-Agent}{Task: Run a symbolic regression
            search for simple single-field slow-roll inflation potentials
        V(phi) that yield \dots.}

        \vspace{0.5em}
        \role{SR Sub-Agent}:
        \action{SearchPotential}
        {Targets: $n_s=0.974 \pm 0.003$, $r=0.0 \pm 0.018$, $N=55$ \\
            Operators: \texttt{[+, -, *, \^{}, /, exp, log, sqrt]} \\
        Evolution: 35 populations $\times$ 45 iterations}
        \annotation{Return the results to the main agent.}

        \vspace{0.5em}
        \role{Main Agent}:
        \action{AnalyzePotential}{Expression: exp(-0.42214/phi)}
        \action{AnalyzePotential}{Expression: exp(-0.00286515*phi**(-1.50769))}
        \dots
    }

    \vspace{0.5em}
    \role{Final Response}:
    Here are two very simple single-field slow-roll potentials that hit
    your target $n_s = 0.974 \pm 0.003$ and also give safely small tensors.
    \begin{itemize}
        \item Candidate A (plateau / inverse-field exponential):
            \[
                V(\phi)=\exp\!\left(-\frac{0.42214}{\phi}\right)
            \]
            Predictions:
            \begin{itemize}
                \item $N=50$: $n_s=0.97271$, $r=0.00564$
                \item $N=55$: $n_s=0.97526$, $r=0.00497$
                \item $N=60$: $n_s=0.97725$, $r=0.00442$
            \end{itemize}
        \item Candidate B (even smaller tensors; ``exp of inverse power'' plateau)
            \[
                V(\phi)=\exp\!\left(-0.00286515\,\phi^{-1.50769}\right)
            \]
            \dots
    \end{itemize}

    \vspace{1em}
    \role{User}: Draw a diagram to analyze the Candidate A.

    \vspace{0.5em}
    \internal{
        \action{PlotPotential}
        Expression: exp(-0.42214/phi)
    }
    \role{Final Response}:
    Saved a 3-panel diagnostic diagram (potential, slow-roll
    parameters, and $n_s$--$r$ overlay) for Candidate A:
    \[
        V(\phi)=\exp\!\left(-\frac{0.42214}{\phi}\right).
    \]
    Plot path: ``\path{candidateA_diagnostics.png}''
\end{agentlog}

The agent successfully discovered a simple potential within a few
minutes, which is consistent with the latest CMB constraints and
predicts a small tensor-to-scalar ratio. In large-$\phi$
approximation, Candidate A corresponds to $V\sim
1-\frac{0.42214}{\phi}$ while Candidate B is $V\sim
1-0.00286515\,\phi^{-3/2}$, both resemble $1-8/\phi^2$ presented
in recent (\citealt{Kallosh:2025rni}) for explaining ACT results.
It is well known that $V\sim 1-{\mu^k/\phi^{k}}$, where $k$ is
positive constant (integer or not), belongs to an extended family
of brane inflation
(\citealt{Kallosh:2019hzo,Kallosh:2018zsi,Dvali:1998pa,Burgess:2001fx,Kachru:2003sx})
and pole inflation
(\citealt{Kallosh:2021mnu,Galante:2014ifa,Terada:2016nqg}).

\subsection{Example: Model Inquiry}

The second example demonstrates the agent's ability to retrieve and
summarize theoretical background from the RAG knowledge base.

Here, we ask the agent to introduce the ``Pseudo-Natural Inflation''
model (\citealt{Arkani-Hamed:2003wrq,Arkani-Hamed:2003xts}) and explain
how it differs from the original ``Natural Inflation''
(\citealt{Freese:1990rb,Adams:1992bn}).
The agent successfully retrieves the relevant section from the
\textit{Encyclop\ae dia Inflationaris}, presents the effective
potential, along with a brief comparison of the two models.

\begin{agentlog}
    \role{User}: Introduce pseudo natural inflation and explain its
    difference with respect to natural inflation.

    \vspace{0.5em}
    \internal{
        \action{Encyclopedia}{query: `Pseudo Natural Inflation model
        potential "pseudo natural inflation"`}
    }

    \vspace{0.5em}
    \role{Final Response}:
    ``\textbf{Pseudo Natural Inflation} (PSNI)'' is an inflation model
    in which the inflaton field is a pseudo-Nambu-Goldstone boson
    associated with the spontaneous breaking of U(1) symmetry. The
    approximate symmetry protects the inflaton potential from large
    radiative corrections (keeping it flat enough for slow roll), while
    a small explicit U(1) breaking generates the tilt of the potential,
    typically via loop effects in a supersymmetric hybrid-type construction.

    A representative effective potential quoted for PSNI is
    \[
        V(\phi)=M^4\left[1+\alpha\ln\left(\cos\frac{\phi}{f}\right)\right],
    \]
    where $M$ sets the overall energy scale by matching the observed
    amplitude of primordial fluctuations, $f$ is the symmetry-breaking
    scale, and $\alpha$ is typically small (often loop-suppressed).
    \vspace{\baselineskip}

    \textbf{How it differs from Natural Inflation (NI):}
    \begin{itemize}
        \item Symmetry origin: NI relies on an (approximate) shift
            symmetry typical of axion-like fields; PSNI is organized around
            a broken U(1) with explicit breaking implemented differently
            (often in SUSY hybrid constructions).
        \item Typical potential: NI gives a cosine potential $V\sim [1\pm
            \cos(\phi/f)]$, whereas PSNI gives a log-cos form $V\sim
            [1+\alpha \ln(\cos(\phi/f))]$ (often loop-induced).
    \end{itemize}

    Source: Encyclop\ae dia Inflationaris (arXiv:1303.3787).
\end{agentlog}

\section{Discussion and Outlook}\label{sec:discussion}

In this work, we have presented \textbf{DeepInflation}, an AI
agent designed for the exploration of inflationary cosmology. By
integrating LLMs with SR and a RAG knowledge base,
\textbf{DeepInflation} automates the discovery of inflationary
potentials that are consistent with observations, while providing
theoretical context grounded in established literature. It
enables researchers and non-experts alike to explore the
inflationary landscape using natural language, thereby
facilitating efficient model discovery and verification.

However, it is important to acknowledge the current limitations of
\textbf{DeepInflation} and outline potential directions for
improvement.

The present framework of \textbf{DeepInflation} focuses on
single-field slow-roll inflation, where
constraints are derived from CMB observables such as $(n_s, r)$.
This approach enables efficient exploration of a broad class of
phenomenologically viable inflationary potentials; however, it
restricts applicability to complex scenarios, such as multi-field
inflation, single-field models with hybrid exit (hybrid inflation
    (\citealt{Linde:1991km,Linde:1993cn}), recent
    e.g.~(\citealt{Kallosh:2022ggf,Ye:2022efx,Braglia:2022phb}) for
implementing $n_s\approx 1$), and null energy condition violating
inflation (\citealt{Piao:2004tq,Liu:2011ns,Cai:2020qpu,Cai:2023uhc}).
Notably, single-field models with an inflection point or a bump in
the potential can lead to enhanced scalar perturbations at small
scales, potentially generating primordial black holes (PBHs), as
discussed in
(\citealt{Garcia-Bellido:2017mdw,Gao:2018pvq,Mishra:2019pzq});
see also (\citealt{Iovino:2025tcv,Allegrini:2025jha}). A critical next step of our
agent project is to extend \textbf{DeepInflation} to discover such
single-field inflationary models, responsible for generating PBHs,
that satisfy both CMB and small-scale constraints.

The discovery of more complex inflationary models requires not
only extended physical calculations but also more robust SR
algorithms. It has been mentioned in
(\citealt{bartlett2025introductionsymbolicregressionphysical})
that since unconstrained SR exploration is prone to combinatorial
explosion, incorporating physically motivated priors into the SR
process is essential. Additionally, integrating LLMs within the SR
loop
(\citealt{shojaee2024llm,grayeli2024symbolicregressionlearnedconcept,harvey2025symbolicregressionmultimodallarge,mesta2024lasr,zhang2025llmmetasrincontextlearningevolving}),
leveraging them for interpreting high-level physical constraints
and modulating the evolutionary search, represents a promising
strategy for guiding the genetic algorithm toward physically
viable regions of the search space prior to numerical
optimization.

The current multi-agent architecture of \textbf{DeepInflation}
comprises a main agent and an SR sub-agent. It is possible that
future iterations will incorporate more specialized agents
dedicated to tasks such as online literature review, data
visualization, and code execution, thereby enhancing the
flexibility and customizability of the current programmatic tools
and enabling the agent to adapt to a wider range of scientific
inquiries. Furthermore, rather than operating as a
human-in-the-loop system that executes tasks explicitly defined by
the user, it is desirable to evolve \textbf{DeepInflation} into a
fully autonomous closed-loop scientific agent that can
independently identify research questions (current constraints on
$n_s-r$, requirement for PBHs, and so on), formulate hypotheses,
and iteratively refine models in the inflationary landscape.

\begin{acknowledgements}
    J.~Z. is supported by the National Natural Science Foundation of
    China (NSFC) under Grants No.~E414660101 and No.~12547104, and the
    Fundamental Research Funds for the Central Universities under
    Grants No.~E4EQ6604X2 and No.~E3ER6601A2. Y.-S.~P. is supported by
    National Key Research and Development Program of China under Grant
    No.2021YFC2203004 and the NSFC under Grant No.12475064.
\end{acknowledgements}

\bibliographystyle{raa}
\bibliography{ref}

\end{document}